\documentclass[journal=JACSAT,manuscript=article]{achemso}
\DeclareUnicodeCharacter{2212}{\textminus}

\usepackage[version=3]{mhchem} 
\usepackage{xcolor}

\author{Jiangnan Liu}
\altaffiliation{These authors contributed equally to this work}

\author{Shuai Liu}
\email{shualiu@umich.edu}
\altaffiliation{These authors contributed equally to this work}

\author{Abdur-Raheem Al-Hallak}
\altaffiliation{These authors contributed equally to this work}

\author{Huabin Yu}

\author{Zhengwei Ye}

\author{Yuheng Zhang}

\author{Zheshen Zhang}
\email{zszh@umich.edu}

\author{Zetian Mi}
\email{ztmi@umich.edu}

\affiliation[UM]
{Department of Electrical Engineering and Computer Science, The University of Michigan, Ann Arbor, Michigan 48109, USA.}

\title[An \textsf{achemso} demo]
  {Hybrid Scandium Aluminum Nitride/Silicon Nitride Integrated Photonic Circuits}

\abbreviations{IR,NMR,UV}

\begin{document}







\begin{abstract}
Scandium doped aluminum nitride has recently emerged as a promising material for quantum photonic integrated circuits (PICs) due to its unique combination of strong second-order nonlinearity, ferroelectricity, piezoelectricity, and  complementary metal–oxide–semiconductor (CMOS) compatibility. However, the relatively high optical loss reported to date —typically above 2.4 dB/cm—remains a key challenge that limits its widespread application in low-loss PICs. Here, we present a monolithically integrated \ce{Si3N4}–ScAlN waveguide platform that overcomes this limitation. By confining light within an etched \ce{Si3N4} waveguide while preserving the functional properties of the underlying ScAlN layer, we achieve an intrinsic quality factor of $Q_{\rm i} = 3.35 \times 10^5$, corresponding to a propagation loss of 1.03 dB/cm—comparable to that of commercial single-mode silicon-on-insulator (SOI) waveguides. This hybrid architecture enables low-loss and scalable fabrication while retaining the advanced functionalities offered by ScAlN, such as the ferroelectricity and piezoelectricity. Our results establish a new pathway for ScAlN-based PICs with potential applications in high-speed optical communication, modulation, sensing, nonlinear optics, and quantum optics within CMOS-compatible platforms.
\end{abstract}

\section{Introduction}
Integrated photonics has been developed as an emerging field to integrate diverse optical and photonic functionalities, including light generation, modulation, sensing, filtering, and computing, onto a compact platform with footprints scaled down to the order of optical wavelengths \cite{SiPICreview, PICreview2, PNN}. The fundamental performance of these integrated photonic devices is critically dictated by the choice of material platform. Silicon nitride (\ce{Si3N4}) has garnered significant attention and developed into a superior candidate, particularly for passive components and nonlinear optical applications that exploit third-order nonlinearity \cite{SiN1,SiN2}. Its compatibility with established CMOS fabrication process enables seamless integration with electronic components while supporting high-volume manufacturing. Furthermore, it exhibits exceptionally low optical loss \cite{SiN422}, boasting intrinsic quality factors ($Q_{\rm i}$) on the order of $10^7$ in microring resonators and propagation losses as low as 0.1$-$1 dB/m across a wide transparency window from the visible to the mid-infrared \cite{SiNfab}. This ultra-low loss is essential for reducing signal attenuation over extended optical paths and fostering high-$Q$ resonant cavities crucial for narrow-linewidth lasers \cite{HighPerfLasersForSiN,HzLineWidthECLSIN}, filters\cite{barwicz2004microring, bryan2023biosensing}, and comb generation \cite{SiNTurnKeySolitonMicroCombs, SiNCombsWaferScale}. However, a significant limitation of \ce{Si3N4} stems from its centrosymmetric crystal structure, which inherently causes the absence of second-order ($\chi^{(2)}$) nonlinear optical effects, preventing efficient electro-optic modulation and second harmonic generation required by active functionalities.

Alternatively, extensive research on lithium niobate (\ce{LiNbO3}) has been conducted for high-performance active photonic components due to its pronounced second-order optical nonlinearity and electro-optic coefficients \cite{LNOchi2, LNOEO}. These properties enable efficient light modulation \cite{TFLNModulatorsReview1, TFLNPhotonicCrystalModulator, TFLNModulatorHighEfficiency2024}, second harmonic generation \cite{TFLNSHGReview, TFLNSHG250K, TFLNSHGDualLayer}, and optical parametric oscillation \cite{TFLNOPANature, TFLNLowThreshOPO}, making it integral to the high-speed modulators in optical communication systems. While traditional bulk \ce{LiNbO3} devices were large and challenging to integrate, recent advancements in thin-film lithium niobate (TFLN) technology have enabled highly confined waveguides and compact devices. However, there are several challenges associated with \ce{LiNbO3}. Direct epitaxial growth of thin TFLN has been limited by the crystalline quality and thickness control. Therefore, most early works on \ce{LiNbO3} devices were built upon bulk substrate, posing challenges for compact integration. Despite techniques such as smart cut that have been developed to produce TFLN, challenges still remain in achieving high-quality and uniform films across a large area \cite{TFLN1}. Additionally, the absence of \ce{LiNbO3}'s direct CMOS compatibility often requires complex heterogeneous integration schemes in the fabrication process. Furthermore, \ce{LiNbO3} typically exhibits higher propagation loss than \ce{Si3N4}, limiting its prospect in loss-sensitive quantum photonics applications. These inherent trade-offs have consistently spurred the development of other material platforms that can offer second-order nonlinear properties and active functionalities while maintaining full CMOS-compatibility and low optical loss.

In this context, aluminum nitride (AlN), a wide bandgap (6.2 eV) nitride semiconductor, has gained increasing interests. AlN is stable in the non-centrosymmetric wurtzite structure and naturally possesses piezoelectric properties and second-order optical nonlinearity \cite{AlNPhotonics1, AlNElctronics}. While primarily known for its applications in deep-UV optoelectronics and high-power electronics, its piezoelectricity and optical properties also position it as a competitive platform for integrated photonics \cite{UltraHighQAlN, SputterAlNEO, AlNCryoComb,QWPockels}. However, the intrinsic second-order nonlinearity of AlN is relatively weak, limiting the efficiency of nonlinear optical processes.

Building upon AlN, scandium aluminum nitride (ScAlN) has emerged as a particularly promising platform for high-performance integrated photonics. Alloying scandium (Sc) into the AlN lattice, as illustrated in Fig. \ref{fig:Materials}(a), offers a unique pathway to significantly enhance its  properties via material engieering~\cite{ScAlNReview,ScAlNPerspective}. ScAlN maintains the wurtzite crystal structure over a wide range of Sc compositions up to about 50\%, before transitioning to a cubic phase \cite{ScAlNTheory}. The incorporation of Sc has led to the emergence of ferroelectricity, and a dramatic increase in both piezoelectricity and, more importantly in terms of photonic applications, the second-order optical nonlinearity \cite{ScAlNFerro, ShubhamPiezoNature, UPenn, TrilayerFerro, AkiyamaPiezo}.
Experimental studies have shown that ScAlN exhibits a second-order nonlinearity an order of magnitude stronger than its pure AlN counterpart. This enhanced property paves a robust path for highly efficient second harmonic generation and cascaded second-order processes for comb generation. Additionally, the emergence of ferroelectricity and greatly enhanced piezoelectricity in ScAlN allows for highly efficient electromechanical transduction, enabling the design of integrated acousto-optic devices for high-speed modulation, frequency shifting, and optical filtering. Recent studies have also suggested that ScAlN can achieve a large electro-optical coefficient, potentially exceeding 50 pm/V at high Sc compositions to surpass that of the conventional \ce{LiNbO3} platform \cite{ScAlNEOHongTang, ScAlNEOTheory}. These outstanding electro-optic and nonlinear optical properties collectively position ScAlN as a superior platform for next-generation high-speed optical modulators, frequency converters, and quantum optical sources to propel optical communication and computing into a new era.

Despite the significant potential of ScAlN, prior efforts in developing passive and active components on this platform have predominantly relied on sputtered ScAlN films. While sputtering offers good deposition rates, it often results in polycrystalline films with higher scattering losses and reduced $Q$ factor \cite{ScAlNSputterMRR}. In this work, we address this limitation by developing a hybrid \ce{Si3N4} and ScAlN platform, wherein single crystalline ScAlN is epitaxially grown by molecular beam epitaxy to achieve lower loss, yielding an intrinsic $Q$-factor of more than $3\times10^5$, significantly higher than previously reported values for monolithic ScAlN platforms \cite{ScAlNSputterMRR, ScAlNMRRNUS}. The meticulously designed hybrid structure aims to combine the ultra-low loss characteristics of \ce{Si3N4} with the powerful nonlinearity and enhanced piezoelectricity provided by ScAlN.

\section{Material Growth and Characterization}
\begin{figure*}[h!]
\centering
\includegraphics[width=1\linewidth]{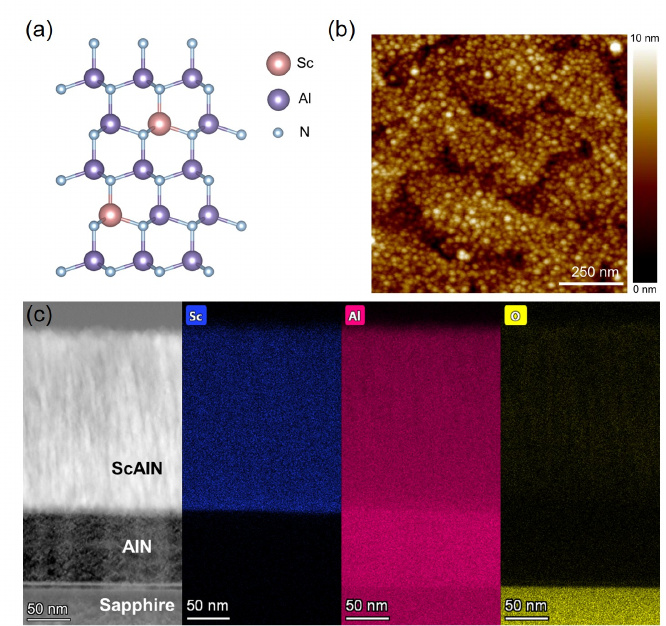}
\caption{{\bf Materials characterization of ScAlN.} (a) Crystal structure illustration of Sc-alloyed AlN. (b) Atomic force microscope imaging of the ScAlN surface showing a roughness of 1.2 nm. (c) Cross-sectional STEM and EDS mappings of the \ce{Si3N4}/ScAlN/AlN/Sapphire structure.}
\label{fig:Materials}
\end{figure*}

The fabrication begins with a sapphire (Al$_2$O$_3$) substrate serving as the bottom cladding layer. A 90 nm AlN layer is grown on the sapphire substrate to facilitate the growth of epitaxial ScAlN, which is critical for minimizing dislocation density and consequently reducing optical loss in the active material. A 200 nm layer of Sc$_{0.1}$Al$_{0.9}$N is then epitaxially grown by plasma-enhanced molecular beam epitaxy (PA-MBE) under ultra-high vacuum, ensuring superior crystallinity and optical properties. Atomic force microscopy characteristic as shown in Fig. \ref{fig:Materials}(b) reveals a surface roughness of 1.2 nm. Further cross-sectional transmission electron microscopy (TEM) and energy-dispersive X-ray spectroscopy (EDX) as shown in Fig. \ref{fig:Materials}(c) clearly shows the scandium incorporation above the AlN buffer layer.

\section{Device Design and Numerical Simulation}

\begin{figure*}[h!]
\centering
\includegraphics[width=1\linewidth]{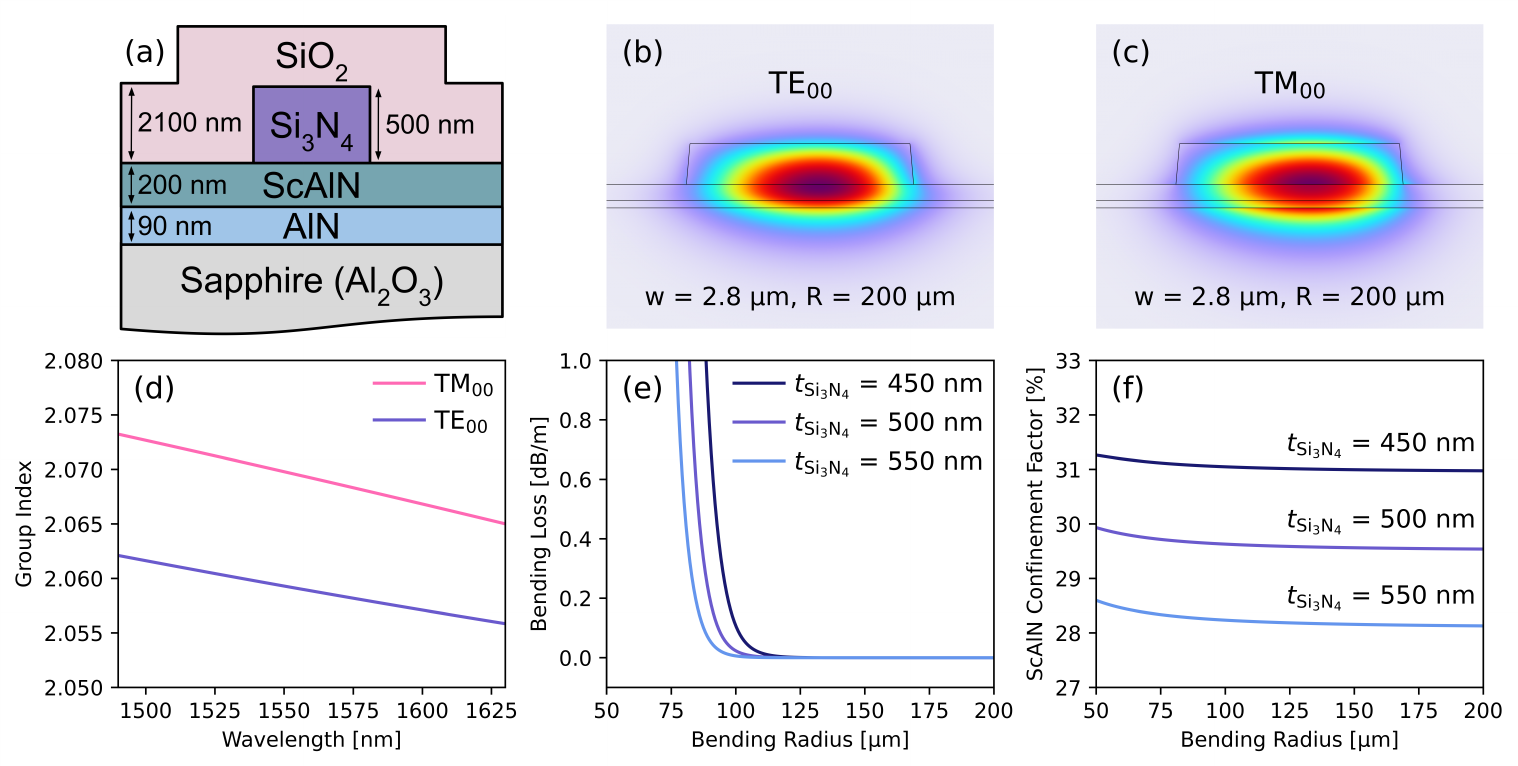}
\caption{{\bf Waveguide design and simulated optical characteristics.} (a) Geometry and dimensions for \ce{Si3N4}-ScAlN hybrid guiding structure. Fundamental TE$_{00}$ (b) and TM$_{00}$ (c) mode distributions ($|\bf{E}|$) in a ring configuration. (d) Group index ($n_g$) for the modes shown in (b) and (c). Bending loss (e) and confinement factor within ScAlN (f) for the TE$_{00}$ mode with respect to bending radius for different thicknesses of \ce{Si3N4} at a guide width of 2.8 $\mu$m.}
\label{fig:Sim}
\end{figure*}

\subsection{Hybrid Waveguide Design Goals}

A major barrier to the widespread adoption of ScAlN is dictated by loss stemming from two sources, material absorption of ScAlN and scattering due to sidewall roughness arising from the etching resistivity. The design presented in this paper, shown in Fig. \ref{fig:Sim}(a), mitigates the first source by controlling the wave's confinement within the ScAlN layer and side-steps the second by leaving the ScAlN un-etched. This design integrates low-loss \ce{Si3N4} with ScAlN and AlN to form a hybrid waveguide that can take advantage of the superior properties of ScAlN while reducing loss, thereby rendering it a promising platform for all-rounder PICs. 

\subsection{Balancing Confinement and Bending Loss}
An important design consideration for the hybrid structure is the ratio of the etched \ce{Si3N4} layer thickness to the thickness of the slab layer of III-N materials. Due to the higher index of ScAlN and AlN in comparison with  \ce{Si3N4}, a significant portion of the light is confined within the slab layer. This is qualitatively manifested by the simulated mode profiles of the fundamental TE and TM modes as shown in Figs. \ref{fig:Sim}(b) and \ref{fig:Sim}(c) respectively. Quantitatively, the confinement factor in ScAlN is plotted in Fig. \ref{fig:Sim}(f), with a comparison of material confinement factors for TE and TM polarization shown in Table \ref{tab:ConfinementFactors}. The group index of the fundamental modes as shown in \ref{fig:Sim}(d), confinement factors plotted in Fig. \ref{fig:Sim} and Table \ref{tab:ConfinementFactors} are determined in COMSOL Multiphysics. The definition of confinement factor is shown in Eq. (\ref{eq:ConfinementFactor}), where $\iint_{\rm X} dA$ represents an integral over a 2D region $\rm X$ and $\iint dA$ represents a 2D integral over all space.

\begin{equation}
    \Gamma_{\rm X} = \frac{\iint_{\rm X} \left | \bf{E} \right|^2 dA}{\iint \left | \bf{E} \right|^2 dA}
    \label{eq:ConfinementFactor}
\end{equation} 
This difference in refractive index allows for a significant confinement factor within ScAlN, despite the ScAlN layer being only less than half as thick as the \ce{Si3N4} layer (200 nm versus 500 nm). Table \ref{tab:ConfinementFactors} shows that the confinement of the fundamental TE mode within the ScAlN layer alone is around 60\% that of the \ce{Si3N4} layer, while its physical thickness is only 40\% of that of \ce{Si3N4}. This confinement is further increased with a reduction in \ce{Si3N4} thickness, albeit coming at the cost of increased bending loss. Since the III-N layer is un-etched, no lateral confinement is present in those layers, allowing for lateral leakage of the mode during strong bending. The interplay between the thickness of \ce{Si3N4}, confinement within the ScAlN layer, and loss is a key parameter of this structure subject to optimization. Some of these dynamics displayed in Figs. \ref{fig:Sim}(e) and \ref{fig:Sim}(f) illustrate the effect of \ce{Si3N4} thickness on bending loss and confinement across a range of bending radii. The fabricated devices discussed in the following sections keep the thickness of \ce{Si3N4} at 500 nm, providing a confinement factor of $\Gamma_{\rm ScAlN}\approx 30\%$ for the fundamental TE mode.

\subsection{Confinement of TE and TM Modes}
Even though the higher index of ScAlN serves to increase confinement within the ScAlN layer for TE modes, a more complex dynamic nonetheless emerge for the TM modes, suggesting reduced confinement within the ScAlN layer for TM modes compared to TE modes, as evidenced in Table \ref{tab:ConfinementFactors}. This is because the sub-wavelength nature of the ScAlN and AlN layers form a light inverse slot mode, which may increase scattering loss from greater interaction with the waveguide boundary while reducing material absorption due to reduced confinement with the layers with higher propagation loss. 

\begin{table}[htbp]
\centering
\caption{Confinement Factors for Fundamental TE and TM Modes$^\textit{a}$}
\begin{tabular}{cccc}
\hline
Polarization & $\Gamma_{\rm Si_3N_4}$ & $\Gamma_{\rm ScAlN}$ & $\Gamma_{\rm AlN}$ \\
\hline
$\rm TE_{00}$ & 47.4\% & 29.5\% & 8.6\% \\
$\rm TM_{00}$ & 45.0\% & 23.0\% & 7.8\% \\
\hline
\end{tabular}
  \label{tab:ConfinementFactors}
  
$^\textit{a}$ For $\lambda = 1550\;{\rm nm},\;w = 2.8\;{\mu \rm m}$
\end{table}

\section{Experimental Results}

\subsection{Device Fabrication}
\begin{figure*}[h!]
\centering
\includegraphics[width=1\linewidth]{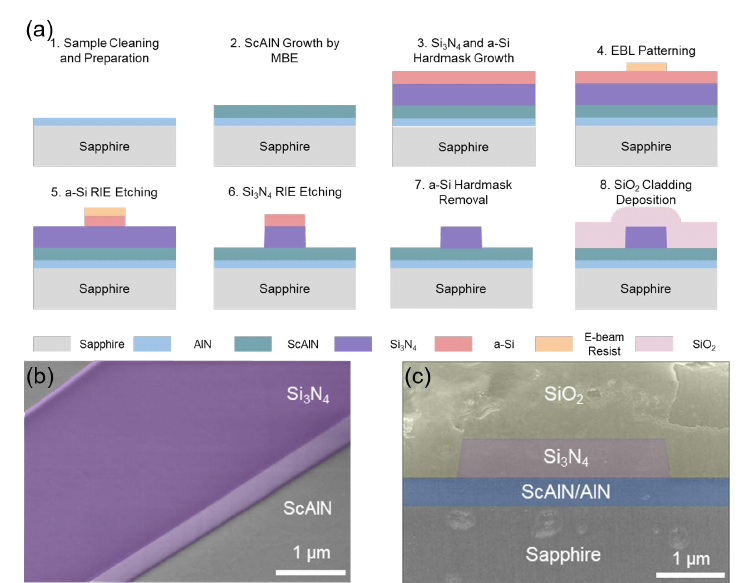}
\caption{ \textbf{Fabrication of the \ce{Si3N4}–ScAlN photonic chips.} (a) Fabrication flow of the monolithically integrated \ce{Si3N4}-ScAlN waveguides. (b) False-colored scanning electron microscopy (SEM) image of the device after \ce{Si3N4} etching showing a smooth sidewall. (c) False-colored cross-sectional SEM of the fabricated waveguide. }
\label{fig:Fab}
\end{figure*}

To fabricate the monolithically integrated \ce{Si3N4}-ScAlN photonic chips, we employ a recently developed amorphous silicon (a-Si) hardmask dry etching technique, enabling high-selectivity and smooth-etched \ce{Si3N4} waveguides\cite{ShuaiACS}. The full fabrication flow is illustrated in Fig. \ref{fig:Fab}(a). The process begins with the MBE-grown ScAlN film on 2-inch single-crystal AlN-on-Al$_2$O$_3$ wafer. On top of this substrate, a 500-nm-thick stoichiometric \ce{Si3N4} layer is deposited via low-pressure chemical vapor deposition (LPCVD), followed by the deposition of a 700-nm-thick a-Si layer as hardmask. Electron beam lithography (EBL) is then performed using MaN-2405 resist, with a beam current of 2 nA, a step size of 8 nm, and exposure dose of 700 $\mu \rm {C/cm^2}$. After resist development, a thermal reflow process is applied on a hotplate to smooth the resist edges and reduce line-edge roughness. The resulting resist pattern is first transferred to the a-Si hardmask via reactive ion etching (RIE) using HBr and He gases. After removing the residual resist, the pattern is etched into the underlying \ce{Si3N4} layer using RIE with a gas mixture of C$_4$F$_8$, CF$_4$, and He. This etch yields waveguides with smooth and nearly vertical sidewalls. The a-Si hardmask is subsequently removed using XeF$_2$ vapor etching, and a SiO$_2$ top cladding is deposited by plasma-enhanced chemical vapor deposition (PECVD). As only the \ce{Si3N4} layer is patterned and etched to define the waveguide, the underlying ScAlN and AlN layers remain intact, avoiding direct etching of these hard-to-etch materials. The scanning electron microscopy (SEM) image in Fig. \ref{fig:Fab}(b) confirms the smooth waveguide sidewalls, indicative of low optical propagation loss. As discussed in the numerical simulations, the rib-like waveguide geometry, shown by the cross-sectional SEM image in Fig. \ref{fig:Fab}(c), enables tight optical mode confinement within the ScAlN layer, while simultaneously benefiting from the low optical loss provided by the high-quality \ce{Si3N4} waveguides.

\subsection{Optical Loss Characterization}
\begin{figure*}[h!]
\centering
\includegraphics[width=1\linewidth]{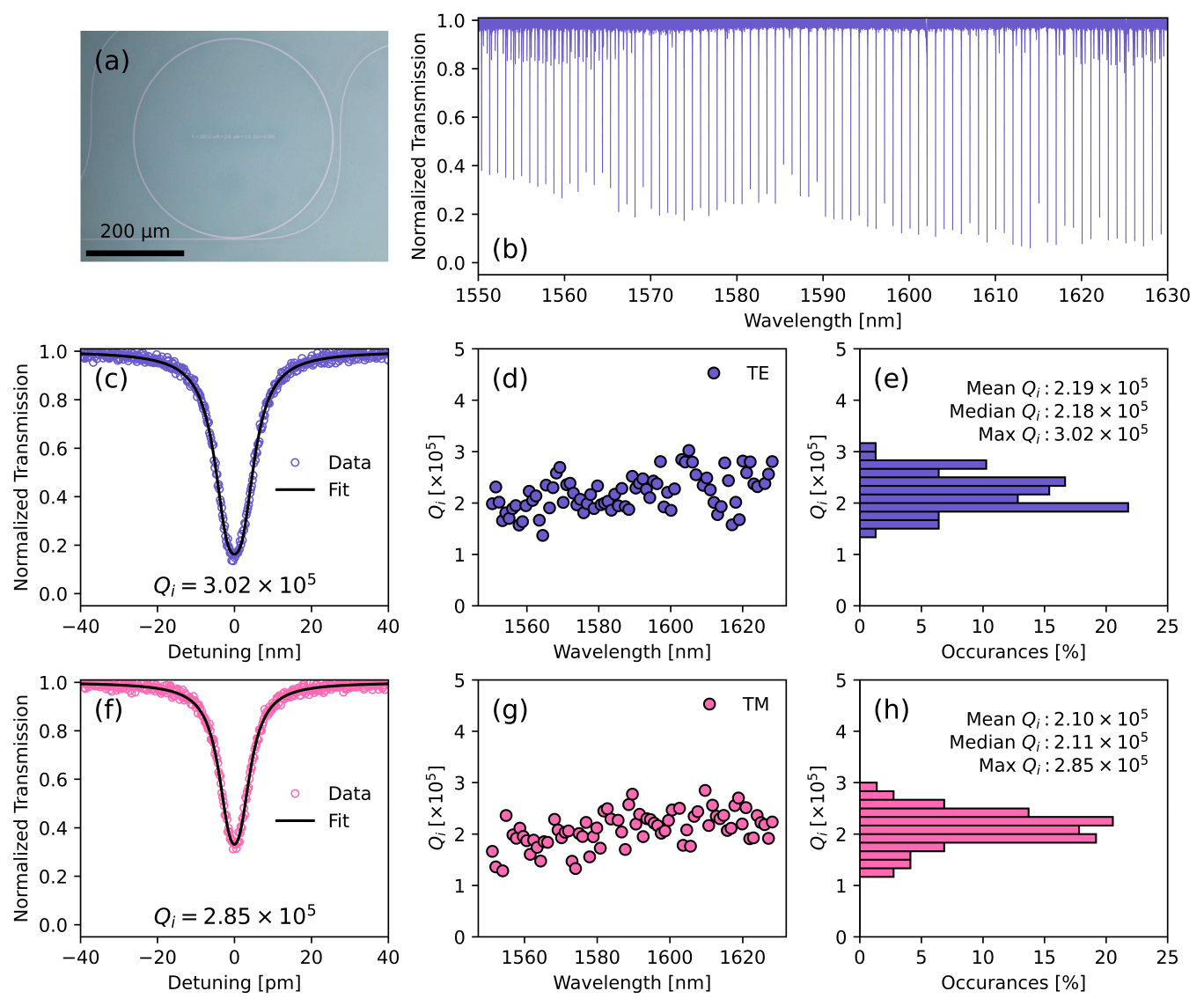}
\caption{ \textbf{Optical loss characterization of the circular microring resonator.} (a) Microscope image of the fabricated \ce{Si3N4}-ScAlN microring resonator ($R$ = 200 $\mu$m, $w_{\rm B}$ = 2.8 $\mu$m). (Scale bar) (b) Normalized transmission spectrum for TE polarization for 1550 to 1630 nm. Fitted highest $Q_{\rm i}$ resonances for TE (c) and TM (f) polarizations, with $Q_{\rm i} = 3.02\times10^5$ and $Q_{\rm i} = 2.85\times10^5$ corresponding to propagation losses of $\alpha = 1.16$ dB/cm and $\alpha=1.23$ dB/cm respectively. Fitted $Q_{\rm i}$ trend with wavelength for TE (d) and TM (g) polarizations. Fitted $Q_{\rm i}$ distributions for TE (e) and TM (h) polarizations.}
\label{fig:MRR}
\end{figure*}

We characterize the propagation loss of the fabricated \ce{Si3N4}-ScAlN microring resonators by measuring their optical transmission spectra using a standard fiber-to-chip edge coupling setup. A narrow-linewidth tunable laser source (TSL-770) is coupled into the chip via a lens fiber and guided into the microring resonator. The transmitted light is then collected by another lens fiber, detected by a low-noise photodetector (Newport 1811), and read out using an oscilloscope. Figure~\ref{fig:MRR}(a) shows the microscope image of a fabricated microring resonator with a radius of $R = 200~\mu\mathrm{m}$ and a waveguide width of $w_{\rm R} = 2.8~\mu\mathrm{m}$. The straight bus waveguide is single-point coupled to the microring with a gap of 500 nm. To ensure good phase matching with the fundamental resonance mode of the microring, the bus waveguide width is set to $w_{\rm B} = 2.0~\mu\mathrm{m}$. The measured normalized transmission spectra for TE modes are presented in Fig.~\ref{fig:MRR}(b). 

We then extract the $Q$ factors by fitting each resonance in the measured transmission spectra, following a similar procedure to that reported in Refs.~\cite{ShuaiACS, PfeifferTS}. We assume that all resonances operate in the under-coupled regime ($Q_{\rm i} < Q_{\rm c}$), where $Q_{\rm i}$ accounts for all intrinsic losses, including material absorption and scattering loss, and $Q_{\rm c}$ denotes the coupling $Q$ factor associated with waveguide-to-resonator coupling loss. Figure~\ref{fig:MRR}(c) shows an example of the fitted high-$Q$ resonance for the TE$_{00}$ mode near $\lambda_0$ = 1605.1 nm, yielding a high $Q_{\rm i}$ = 3.02 $\times~10^5$. The corresponding propagation loss is calculated using Eq. (\ref{eq:loss}), 
\begin{equation}
    \alpha = \frac{2\pi n_g}{Q_{\rm i} \lambda_0},
    \label{eq:loss}
\end{equation}
where the group index is $n_{\rm g} = 2.06$, resulting in a propagation loss of $\alpha = 1.16$ dB/cm. Figure~\ref{fig:MRR}(d) summarizes all extracted $Q_{\rm i}$ values across the 1550–1630 nm wavelength range, each fitted with a goodness-of-fit (R-square) exceeding 95\% \cite{ShuaiACS}. The statistical distribution of the extracted $Q_{\rm i}$ factors is presented in Fig.~\ref{fig:MRR}(e), showing a mean high $Q_{\rm i} = 2.19 \times 10^5$. For TM modes, the max $Q_{\rm i}$ reaches $2.85 \times 10^5$ (in Fig.~\ref{fig:MRR}f) near $\lambda_0 = 1609.7$ nm, corresponding to a propagation loss of $\alpha = 1.23$ dB/cm. Figs.~\ref{fig:MRR}(g) and \ref{fig:MRR}(f) show the corresponding extracted $Q_{\rm i}$ values and their statistical distribution, with a mean $Q_{\rm i} = 2.10 \times 10^5$. Compared with the previously reported maximum $Q_{\rm i}$ of approximately $1.5 \times 10^5$ \cite{DiZhuHighQ, ScAlNEOHongTang} for pure ScAlN waveguides, our consistently high-$Q$ factors highlight the robustness of monolithic \ce{Si3N4}-ScAlN design and the fabrication process, enabled by both the high crystalline quality of the MBE-grown ScAlN film and the low-loss etching of the \ce{Si3N4} waveguides.  

\begin{figure*}[h!]
\centering
\includegraphics[width=1\linewidth]{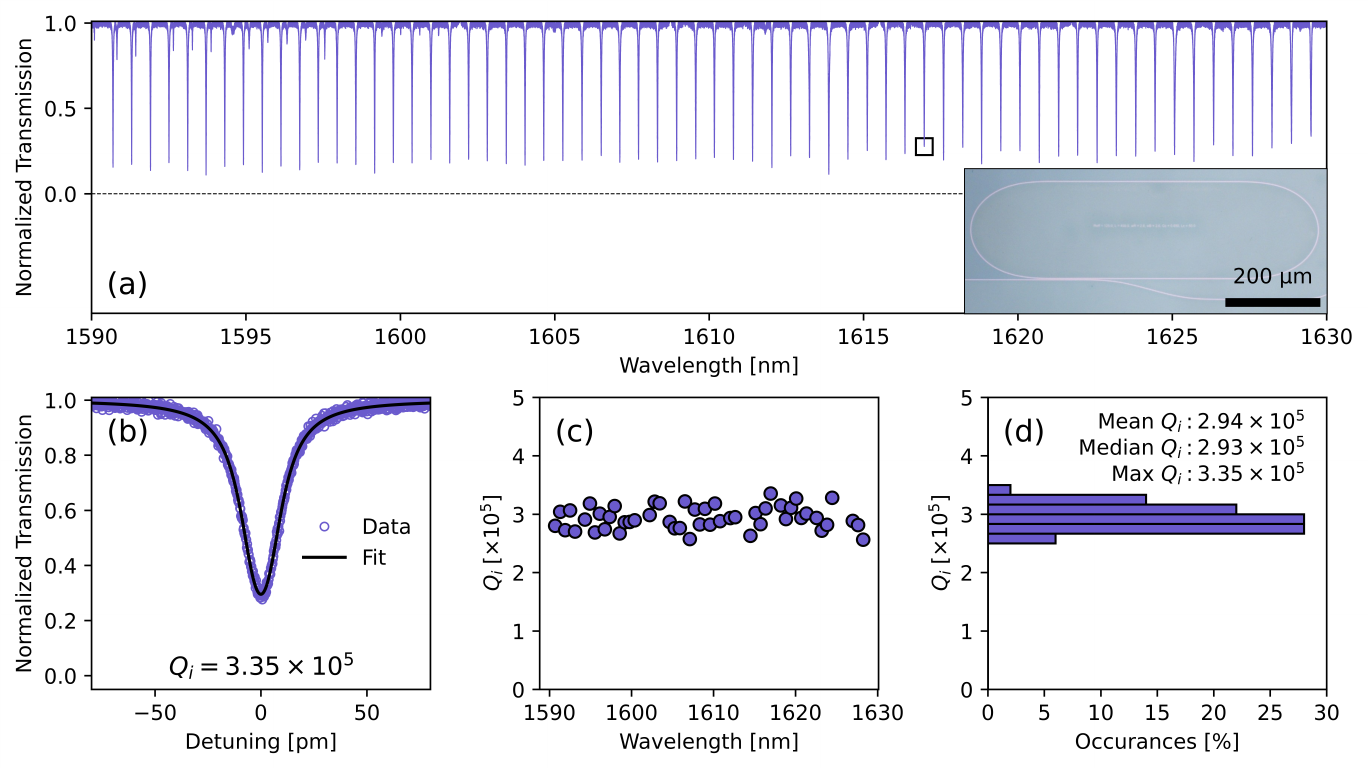}
\caption{\textbf{Optical loss characterization of the Euler racetrack microring resonator.} (a) Normalized transmission spectrum from 1590 nm to 1630 nm for TE polarization, with highest $Q_{\rm i}$ resonance highlighted and Euler-bend racetrack resonator ($R_{\rm eff}$ = 125 $\mu$m, $w$ = 2.6 $\mu$m) inset. (b) Fitted highest intrinsic quality factor resonance for TE polarization, with $Q_{\rm i} = 3.35\times10^5$ corresponding to a propagation loss of $\alpha$ = 1.03 dB/cm. Dots: experiment data. Solid line: fitting.  (c) Fitted $Q_{\rm i}$ trend with wavelength for TE polarization. (d) Fitted $Q_{\rm i}$ distribution for TE polarization.}
\label{fig:Euler}
\end{figure*}

To further reduce optical loss in the microring resonator, we implement an Euler racetrack design consisting of two straight waveguide sections connected by 180$^\circ$ adiabatic Euler bends at both ends~\cite{SiEulerRacetrack, ji2022compact}. The straight waveguide segments confine the optical mode in the waveguide center, thereby minimizing interaction with the sidewalls and reducing scattering loss compared to conventional circular microrings. The Euler bends provide an adiabatic curvature transition between the straight and curved sections to help suppress the mode mismatch induced coupling from fundamental to higher-order modes typically associated with larger optical loss. An optical microscope image of the fabricated Euler racetrack microresonator is shown in the inset of Fig.~\ref{fig:Euler}(a). The effective bending radius of the Euler curves is $R_{\rm eff} = 125~\mu\rm{m}$, offering a balance between negligible bending loss (as shown in Fig.~\ref{fig:Sim}(e)) and a compact footprint. We design the straight waveguide sections with a length of $L_s = 450~\mu\rm{m}$, resulting in a total cavity length of $L = 2041~\mu\rm{m}$. The bus and resonator waveguides share identical width ($w_{\rm B} = w_{\rm R} = 2.6~\mu\rm{m}$) and remain straight in the coupling region to naturally satisfy the perfect phase-matching condition over a broad wavelength range. Together with the adiabatic Euler bends, this configuration enables efficient excitation and low-loss propagation of the fundamental modes within the resonator. The bus-to-resonator gap is $g = 600~\rm{nm}$ with a coupling length of $L_{\rm c} = 50~\mu\rm{m}$. 

Figure~\ref{fig:Euler}(a) shows the normalized transmission spectrum of the TE modes in the Euler racetrack microresonator operating in the slightly over-coupled regime ($Q_{\rm i} > Q_{\rm c}$). Owing to the incorporation of long straight waveguide sections, the max intrinsic quality factor improves to $Q_{\rm i} = 3.35 \times 10^5$, as evidenced by the fitted resonance shown in Fig.~\ref{fig:Euler}(b). At the resonance wavelength of $\lambda_0 = 1616.97$ nm with a group index of $n_{\rm g} = 2.05$, the calculated propagation loss $\alpha = 1.03~\rm{dB/cm}$. This low propagation loss compares favorably with the 1–2 dB/cm of single-mode silicon waveguides on the well-established commercial 220-nm silicon-on-insulator (SOI) platform \cite{Si2p4dBpercm, SiBeyondSM} that have enabled some of the most advanced PICs for emerging applications in high-speed data centers and optical communications. The demonstrated performance of our hybrid \ce{Si3N4}-ScAlN platform thus offers competitive low-loss characteristics suitable for next-generation integrated photonics, while also featuring additional functionalities including high $\chi^{(2)}$ nonlinearity, ferroelectricity, and piezoelectricity. It is also evident from Figs.~\ref{fig:Euler}(c) and \ref{fig:Euler}(d) that this high quality factor is not an outlier, as corroborated by a tight distribution of high intrinsic quality factors that exhibits a large mean value of $Q_{\rm i} = 2.94 \times10^5$, indicating the robustness of the low-loss performance of these \ce{Si3N4}-ScAlN PICs.


\section{Conclusion}
In summary, we experimentally demonstrate a monolithically integrated \ce{Si3N4}–ScAlN photonic platform with an intrinsic quality factor of $Q_{\rm i} = 3.35 \times 10^5$, corresponding to a low propagation loss of 1.03 dB/cm on par with hat of commercial single-mode SOI waveguides. The optical mode is confined within the etched \ce{Si3N4} layer, avoiding direct etching of the hard-to-process underlying ScAlN film. This monolithically integrated structure benefits from the mature, low-loss fabrication process of \ce{Si3N4} waveguides while leveraging the functional properties of the ScAlN layer. It is also worth noting that further reduction in optical loss is foreseeable as the current \ce{Si3N4} film's residual absorption due to hydrogen-related bonds (e.g., N–H, Si–H) can be mitigated through proper annealing or hydrogen-free deposition techniques toward even lower material absorption and improved waveguide performance \cite{bose2024anneal,zhang2024low}. Our results highlight the potential of combining multiple functional materials and components on a single chip, enabling advanced photonic functionalities while remaining compatible with the well-established CMOS fabrication processes to pave the way for scalable, high-speed photonic systems tailored to future data communication, signal processing, and AI-driven big data center applications.

\begin{acknowledgement}
JL, AA, HY, ZY and ZM acknowledge Army Research Office under Grant No.~W911NF2420210 and National Science Foundation under Grant No.~2435166. SL and ZZ acknowledge National Science Foundation under Grant No.~2326780, No.~2330310, and No.~2317471. Authors also acknowledge technical support from the Lurie Nanofabrication Facility (LNF) at the University of Michigan.
\end{acknowledgement}

\section{Data availability}
Data underlying the results presented in this paper are not publicly available at this time but may be obtained from the authors upon reasonable request.

\section{Disclosures}
The authors declare no conflicts of interest.

\bibliography{achemso-demo}

\end{document}